%% file: main.tex
\documentclass[sigconf]{acmart}

\usepackage{listings}
\usepackage{subcaption}
\usepackage{acro}
\usepackage{calc}
\usepackage{mathpartir}
\usepackage{mathtools}

\AtBeginDocument{%
  \providecommand\BibTeX{{%
    \normalfont B\kern-0.5em{\scshape i\kern-0.25em b}\kern-0.8em\TeX}}}

\setcopyright{acmcopyright}
\copyrightyear{2021}
\acmYear{2021}

\acmConference[ICOOOLPS '21]{ICOOOLPS '21}{July 13, 2021}{Online}

\lstdefinelanguage{fuel}{
keywords={%
  unit, junk, true, false, nil, func, salloc, halloc, at, store, load, call,%
  if, else, while, assuming, unwrapping, @own, @brw, @dyn%
},%
escapeinside={(*}{*)},%
morecomment = [l]{//},%
}

\lstdefinelanguage{rust}{
keywords={%
  true, false, unsafe, async, await, move, use, pub, crate, super, self, mod, struct, enum, fn, const, static, let, mut, ref, type, impl, dyn, trait, where, as, break, continue, if, else, while, for, loop, match, return, yield, in
},%
ndkeywords={
        bool,u8,u16,u32,u64,u128,i8,i16,i32,i64,i128,char,str,
        Self,Option,Some,None,Result,Ok,Err,String,Box,Vec,Rc,Arc,Cell,RefCell,HashMap,BTreeMap,
        macro_rules
    },
escapeinside={(*}{*)},%
morecomment = [l]{//},%
}

\begin{document}

\title{Fuel: A Compiler Framework for Safe Memory Management}

\author{Dimitri Racordon}
\email{dimitri.racordon@unige.ch}
\affiliation{%
  \institution{University of Geneva}
  \country{Switzerland}
}

\author{Aurélien Coet}
\email{aurelien.coet@unige.ch}
\affiliation{%
  \institution{University of Geneva}
  \country{Switzerland}
}

\author{Didier Buchs}
\email{didier.buchs@unige.ch}
\affiliation{%
  \institution{University of Geneva}
  \country{Switzerland}
}

\renewcommand{\shortauthors}{Racordon et al.}

\begin{abstract}
  Flow-sensitive type systems offer an elegant way to ensure memory-safety in programming languages.
  Unfortunately, their adoption in new or existing languages is often hindered by a painful effort to implement or integrate them into compilers.
  This paper presents early results in our effort to alleviate this task.
  We introduce Fuel, a type capability-based library that can be plugged onto a compiler toolchain to check for memory-safety properties.
  Fuel builds upon well-established ideas in the domain of capability-based system, and adds a mechanism leveraging dynamic checks to recover capabilities where static reasoning is either too difficult or impossible.
  This approach allows the analysis to potentially cover situations where a typical type system might not be expressive enough to statically reason about memory safety.
\end{abstract}



\maketitle

\input{introduction}
\input{relatedwork}
\input{fuel}
\input{conclusion}

\bibliographystyle{ACM-Reference-Format}
\bibliography{main}

\end{document}

%% file: introduction.tex
\section{Introduction}

Memory-related errors are a pernicious source of bugs in software~\cite{DBLP:journals/ieeesp/SzekeresPWS14}.
Although they are most commonly associated to programming languages with explicit memory management, even so-called ``managed'' languages that hide memory handling operations from the user are not completely immune to the problem~\cite{DBLP:journals/pacmpl/KabirLL20}.
As a consequence, significant research effort has been put into finding ways to alleviate memory-related bugs, resulting in a wide range of static and dynamic analyses.

Recent years, in particular, have seen a surge of interest in advanced type systems that can be used to statically prove memory safety properties.
Type systems not only offer an elegant way of classifying values and the operations allowed on them, they are also one of the most scalable tools to formally reason about the correctness of programs.
Traditional type systems, however, are a poor fit for memory safety checking, because they struggle to accurately track properties of shared mutable state.
As a countermeasure, \emph{flow-sensitive} type systems~\cite{ACM:journal/sigplan/Foster02} have been developed.
Flow-sensitive type system combine traditional type-checking with data flow analysis to constrain aliasing and keep track of properties that depend on a program’s execution flow.
Hence, these systems are particularly well-suited for the detection of memory-related bugs.

Unfortunately, although flow-sensitive approaches have been successfully implemented in industrial-strength languages like \emph{Rust} or \emph{Pony}, their adoption in other new or existing languages is often met with a surprisingly painful implementation effort.
In many cases, novel language features can be ``encoded'' within an existing host language, using clever meta-programming tricks~\cite{DBLP:journals/pacmpl/BallantyneKF20}, at a relatively low engineering cost.
However, this approach is delicate to adopt when extensions require a more intimate understanding of the program's semantics.

This paper presents early results in our effort to ease the implementation burden of type-based memory safety checking.
We introduce \emph{Fuel}, a framework that can be plugged into a compiler's pipeline to check programs against memory-related bugs.
Fuel is centered around a low-level language, reminiscent of LLVM IR, which relies on type capabilities to keep track of mutable states and describe aliasing relationships.
Fuel supports intra-procedural reasoning, using annotations to thread typing assumptions across function boundaries.
Further, it advocates for the use of dynamic checks to recover static assumptions in places where static reasoning is either impractical or impossible.

%% file: relatedwork.tex
\section{Related Work}

This paper defines ``memory safety'' in terms of properties ensuring that accesses to a machine’s memory do not cause crashes or undefined behavior.
More specifically, we identify three main categories of memory errors:
\begin{description}
  \item[Invalid dereference errors] These occur when a program attempts to read at a location that is no longer allocated (a.k.a. use-after-free) or that does not exist (e.g., \texttt{null} dereference).
  \item[Invalid deallocation errors] These occur when a program attempts to deallocate memory that was already freed, or that must not be freed manually.
  \item[Memory leaks] These occur when a program does not deallocate memory that is no longer needed or accessible.
\end{description}

\paragraph{Static analysis}

Over the years, several forms of static analysis have been proposed to eliminate memory management errors.
A method that has seen considerable success in proving properties on programs while remaining both scalable and modular is \emph{type checking}.
Good representative of this approach include linear~\cite{DBLP:conf/ifip2/Wadler90} and affine~\cite{DBLP:conf/popl/TovP11} type systems, as well as ownership types~\cite{DBLP:series/lncs/ClarkeOSW13} and, more closely related to our work, capability-based systems~\cite{DBLP:conf/ecoop/BoylandNR01}.

In a capability-based system, types are extended with context-sensitive properties describing which operations are allowed on program variables with regard to memory management.
Capabilities generally aim at formalizing \emph{uniqueness} and \emph{immutability}.
Both serve to limit unintended mutations and to prove data-race freedom in concurrent settings.
One advantage of this technique is that it dissociates pointer values, which can be copied freely, from the permission to use them~\cite{DBLP:conf/esop/SmithWM00}, making the representation of mutable self-referential data structures possible.

\paragraph{Dynamic analysis}

Other approaches, such as dynamic binary instrumentation~\cite{DBLP:journals/ijhpca/BuckH00} and compile time instrumentation~\cite{DBLP:conf/cgo/StepanovS15} can be used to fend against memory bugs.
Both forms of analysis instrument the code of a program to monitor it and detect memory-related errors during its execution.

In principle, dynamic analyses are more precise than static ones, as they can observe the actual program  without relying on conservative assumption about its runtime behavior.
Further, dynamic analyses are often language-agnostic, as they tend to operate on native or object code rather than actual language sources.

%% file: fuel.tex
\section{The Fuel Framework}




Fuel is a library that can be plugged into existing compilers to check for memory-safety properties.
Although it is still under active development, a functional implementation is already available as an open-source project\footnote{
https://github.com/kyouko-taiga/fuel
}.

At the center of the library is an intermediate language (IL) that resembles a RISC-like instruction set in which memory traffic is expressed explicitly.
The IL is equipped with type capabilities to keep track of ownership and permissions, and it offers runtime checks on type assumptions as an escape hatch when static reasoning is not possible.
The remainder of this section offers a quick tour Fuel's language through small examples of C and Rust programs translated to the IL and checked for memory-related bugs.

\subsection{Initialization Tracking}

Figure \ref{fig:var-init-rust} presents a small Rust program.
Two variables \lstinline|b| and \lstinline|i| are declared, of types \lstinline|bool| and \lstinline|i32| (32-bit signed integer), respectively.
\lstinline|b| is assigned and its value is used to determine \lstinline|i|'s.

The Rust program is translated to Fuel's IL in Figure \ref{fig:var-init-fuel}.
Instructions of the form \lstinline|x = instruction| declare local, temporary registers that cannot be reassigned, similar to variables in \emph{static single assignment} (SSA) form~\cite{DBLP:journals/toplas/CytronFRWZ91}.
Those registers designate immutable values that are valid for the duration of their scope.
The two variables from the Rust program are translated to allocations, each resulting in the creation of a \emph{memory cell} whose address is assigned to a register.
The notation \lstinline|at m0| on the first line gives the name \lstinline|m0| to the cell being allocated.

\begin{figure}[ht]
%
%
\begin{minipage}[t]{.49\linewidth}
\begin{lstlisting}[language=rust]
let b: bool;
let i: i32;
b = true;

if b { i = 2; }
else { i = 4; }
\end{lstlisting}
\vfill
\subcaption{Rust implementation}
\label{fig:var-init-rust}
\end{minipage}
\hfill
%
%
\begin{minipage}[t]{.50\linewidth}
\begin{lstlisting}[language=fuel]
breg = salloc Bool at m0
ireg = salloc I32 at m1
store true, breg
bval = load breg
if bval { store 2, ireg }
else    { store 4, ireg }
\end{lstlisting}
\subcaption{Equivalent code in Fuel IR}
\label{fig:var-init-fuel}
\end{minipage}
%
%
\caption{Declaration and initialization of local variables.}
\label{fig:var-init}
\end{figure}

In Fuel, each register declaration creates a new type assumption describing the type of data stored in the register.
We refer to these assumptions as \emph{register capabilities}, because they describe the operations that are supported on the register's value.
The stack allocation at line 1 produces such a capability.
However, unlike Rust, Fuel does not map \lstinline|breg| to \lstinline|Bool|.
Instead, it creates a capability \lstinline|[breg: !m0]|, where \lstinline|!m0| designates a singleton type~\cite{DBLP:conf/csl/Aspinall94} whose unique inhabitant is the address of a memory cell named \lstinline|m0|.

The allocation also creates a \emph{cell capability}, denoted here by \lstinline|[m0: Junk<Bool>]|, capturing the intuition that, although the cell was declared containing boolean values, it currently contains uninitialized garbage.
At line 3, the store instruction \emph{consumes} \lstinline|[m0: Junk<Bool>]| to \emph{produce} a new capability \lstinline|[m0: Bool]|, effectively resulting in a strong update~\cite{DBLP:conf/esop/SmithWM00} and moving the cell's state to \emph{initialized}.

If \lstinline|b|'s assignment at line 3 was delayed until after the conditional statement (e.g., at line 7), the program would use the value of an uninitialized boolean and hence expose undefined behavior.
However, thanks to Fuel's type system, in the absence of the store instruction at line 3, the capability for the cell \lstinline|m0| remains unchanged, which makes the program ill-typed.

\subsection{Alias Tracking}

Fuel represents memory locations at the type system's level.
Hence, tracking aliases is straightforward.
Consider the C program in Figure \ref{fig:deref-c}.
It declares a variable \lstinline|foo| and assigns its address to pointer \lstinline|bar| at line 3.
Since \lstinline|foo|'s value isn't initialized until line 4, \lstinline|bar| points to uninitialized memory after its assignment.
The last line dereferences \lstinline|bar|, increments its value and assigns it to \lstinline|foo|.

\begin{figure}[ht]
%
%
\begin{minipage}[b]{.49\linewidth}
\begin{lstlisting}[language=c]
float  foo;
float* bar;
bar = &foo;
foo = 13.37f;

foo = 1.0f + *bar;
\end{lstlisting}
\subcaption{C implementation}
\label{fig:deref-c}
\end{minipage}
\hfill
%
%
\begin{minipage}[b]{.49\linewidth}
\begin{lstlisting}[language=fuel]
foo = salloc F32 at m0
bar = salloc (*$\exists$*)a.!a at m1
store foo, bar
store 13.37f, foo
t0 = load bar
t1 = load t0
t2 = call add, 1.0f, t1
store t2, foo
\end{lstlisting}
\subcaption{Equivalent code in Fuel IR}
\label{fig:deref-fuel}
\end{minipage}
%
%
\caption{Pointer dereference.}
\label{fig:deref}
\end{figure}

The translation to Fuel relies on slightly more involved concepts.
In particular, the second stack allocation uses an existential type~\cite[Chapter~24]{DBLP:books/daglib/0005958} \lstinline[mathescape]|$\exists$a.!a| to represent the pointer type \lstinline[language=c]|float*| from the C program.
Intuitively, an existential type captures the idea that some specific details about the actual type are hidden.
In this case, the specific cell \lstinline|a| whose address will be stored in \lstinline|bar| is existentially quantified because it is unknown when the register is declared.
Remark also that \lstinline[mathescape]|$\exists$a.!a| is not as precise as \lstinline[language=c]|float*|, since the address alone does not specify the kind of values to which it refers.
That is because a stack allocation of the form \lstinline[mathescape]|x = salloc $\tau$ at m| uses the type argument only to specify the layout of the values stored in \lstinline|m|.
Since the layout of an address does not depend on the values it points to (it is simply a pointer), any address type is valid.

As in our first example, the stack allocations at lines 1 and 2 produce two capabilities.
The second maps \lstinline|m1| to \lstinline[mathescape]|Junk<$\exists$a.!a>| and is consumed at line 3 to strongly update \lstinline|m1|, resulting in \lstinline|[m1: !m0]|.
Two load instructions translate the \lstinline|*bar| expression on the last line of the C program.
The first, at line 5, loads the contents of the cell \lstinline|m1|, that is, the address of \lstinline|m0|, into register \lstinline|t0|.
The instruction results in the creation of a capability \lstinline|[t0: !m0]|, revealing that \lstinline|t0| has an address type.
The second load, at line 6, fetches the contents of cell \lstinline|m0|, the actual floating point number, into register \lstinline|t1|, resulting in the creation of a capability \lstinline|[t1: F32]|.
The remainder of the program is straightforward: the call instruction applies the function \lstinline|add|, whose result is eventually stored into \lstinline|m0|.

The C program from Figure \ref{fig:deref-c} would exhibit undefined behavior if \lstinline|foo|'s initial assignment was omitted.
Remark, however, that the issue would not be caused by the assignment at line 3, since taking the address of an uninitialized variable is not inherently wrong.
The problem would occur at line 8, when \lstinline|bar| is dereferenced.
Similar to our previous example, this last instruction would not type check, as it would attempt to dereference a junk value.

\subsection{Function Boundaries and Borrowing}

Intra-procedural approaches to type checking can abstract away portions of the code (e.g., when the actual source code of a function is unavailable), which allows modular analyses.
This advantage comes at the cost of requiring that function signatures serve as complete interfaces, describing how data flows in and out.
Reference semantics complicate this requirement because of possible side-effects: if a function modifies the contents of a cell referred to by one of its arguments, then one cannot rely on its co-domain alone to determine the result of its application.
For instance, consider the function header \lstinline[language=c]{int libfoo_f(int*)} in C: the function might simply consult the value pointed by its argument, causing no side effect, but it may also modify it or even free the pointer.
To address this issue, Fuel allows capabilities to occur in domains and/or codomains.
In other words, a function's signature not only describes the type of its parameters, but also captures pre- and post-conditions about its caller's typing environment.

An example of a Rust program that uses an externally defined function is proposed in Figure \ref{fig:rust_fn_brw}.
The implementation of the function \lstinline{libfoo_f} is unknown, meaning that we cannot analyze its behavior to deduce possible side effects and can only rely on its signature for type checking.
Fortunately, the signature describes how the function's parameters might be affected during a call.
The first parameter is \emph{borrowed mutably}, which indicates that \lstinline{libfoo_f} can mutate it but must return the full capability for the argument to the caller after it returns.
Parameter \lstinline|b| is borrowed \emph{immutably}, meaning that the function only has read access to it.

\begin{figure}[ht]
\centering
\begin{lstlisting}[language=rust]
fn libfoo_f(a: &mut i32, b: &i32) -> i32

fn main() {
  let mut a = 1;
  let b = 42;
  a = libfoo_f(&mut a, &b);
}
\end{lstlisting}
\caption{Example of a Rust program using an externally defined function borrowing its parameters}
\label{fig:rust_fn_brw}
\end{figure}

The translation of the Rust program to Fuel's IL is given in figure \ref{fig:fuel_fn_brw}.
Function \lstinline{libfoo_f}'s signature is expressed with a \emph{universal type}~\cite[Chapter~23]{DBLP:books/daglib/0005958} quantifying \lstinline|a| and \lstinline|b| over all possible memory cells.
Polymorphism is necessary to abstract over the actual name of the cell with which the function will be called~\cite{DBLP:journals/lisp/TofteBEH04}.
The notation \lstinline[language=fuel]{(!a,!b)+[a: I32, @brw(b: I32)]} reads as ``the singleton types \lstinline|!a| and \lstinline|!b| and the assumption that the cells named \lstinline|a| and \lstinline|b| contain integer values''.
The annotation \lstinline[language=fuel]{@brw} further indicates that the capability for \lstinline|b| is \emph{non-linear} (or \emph{borrowed}): it can be copied at will and used to read the cell's contents, but it cannot mutate or deallocate it.
Borrowed capabilities can only be obtained by weakening linear capabilities at function boundaries.
This mechanism naturally delineates the duration of the loan, which expires when the function returns.

\begin{figure}[ht]
\centering
\begin{lstlisting}[language=fuel]
func libfoo_f(_0, _1):
  (*$\forall$*)a,b.(!a,!b)+[a: I32, @brw(b: I32)]
  -> I32+[a: I32]

func main(): () -> () {
  a = salloc I32 at m0
  store 1, a
  b = salloc I32 at m1
  store 42, b
  v = call libfoo_f, a, b
  store v, a
}
\end{lstlisting}
\caption{Translation of the program from figure \ref{fig:rust_fn_brw} to Fuel}
\label{fig:fuel_fn_brw}
\end{figure}

The mutable borrow of \lstinline|a| is represented in Fuel by the consumption of the parameter's capability by \lstinline{libfoo_f}, followed by its restitution to the caller when it returns.
The notation \lstinline[language=fuel]{I32+[a: I32]} indicates that \lstinline|a|'s capability is preserved by \lstinline{libfoo_f}'s return type: after a call to it, the type system guarantees that the cell still contains a valid 32-bit signed integer.
Any failure by the function to preserve this capability would result in an ill-typed program, and therefore be statically rejected by the framework.

\subsection{Runtime Checks}

One weakness of the type system we have presented so far is that it does not allow a function to conditionally consume or produce linear capabilities.
Consider the C function \lstinline[language=c]{void free_one(int* x, int* y)} that would accept two pointers and free only one of them, depending on some condition.
This function would necessarily need to consume linear capabilities for both its parameters to be able to free either one of them.
A defensive strategy would be to assume that both arguments are freed by the function when it is called.
This approach would however inevitably lead to a memory leak, as the call site would definitely lose the ability to deallocate the memory referenced by the pointer left intact by the function.

Fuel addresses this limitation with dynamic checks to test for specific type capabilities at runtime.
Just as borrowed capabilities, capabilities qualified by \lstinline|@dyn| (for \emph{dynamic}) are also treated non-linearly.
Unlike borrows, however, they can be used to perform mutating operations, provided their use is guarded by a runtime check that the associated cell can be safely dereferenced\footnote{The framework requires that every use of a dynamic capability be guarded.}.

\begin{figure}[ht]
\begin{lstlisting}[language=fuel]
func free_one(x, y):
  (*$\forall$*)a,b.(!a,!b)+[a:@dyn(I32), b:@dyn(I32)] -> Void

func main(): () -> () {
  i = halloc I32 at m0       // => [m0: Junk<I32>]
  j = halloc I32 at m1       // => [m1: Junk<I32>]
  store 42, i                // => [m0: I32]
  store 24, j                // => [m1: I32]
  _ = call free_one, i, j    // trades [m0: I32,
                             //         m1: I32]
                             // for [@dyn(m0:I32),
                             //      @dyn(m1:I32)]
  assuming i: i32 { free i } // recover [m0: I32]
                             // to free m0
  assuming j: i32 { free j } // recover [m1: I32]
                             // to free m1
}
\end{lstlisting}
\caption{A program using dynamic capabilities.}
\label{fig:dynamic-assumptions}
\end{figure}

Figure \ref{fig:dynamic-assumptions} showcases dynamic capabilities.
Function \lstinline{free_one}'s signature is defined so that it accepts two dynamic capabilities for its parameters.
At the function's call site, at line 9, the capabilities for \lstinline{m0} and \lstinline{m1} are weakened to \lstinline[language=fuel]{[m0: @dyn(I32)]} and \lstinline[language=fuel]{[m1: @dyn(I32)]}, effectively making them dynamic and non-linear.
After the call, neither of these capabilities can be accessed directly to read or write the memory to which they are associated.
Instead, they must be guarded by dynamic checks, as illustrated by the last two lines of the example.
The statement \lstinline[language=fuel]{assuming i: i32} tests whether \lstinline|[@dyn(m0: I32)]| can be traded for a regular capability \lstinline|[m0: I32]| at runtime, in which case the command \lstinline[language=fuel]{free i} is executed, consuming \lstinline|[m0: I32]| to delete the associated memory cell.

An important restriction that must be guaranteed at runtime to preserve the soundness of this mechanism is that a dynamic capability cannot be traded for a regular one if another capability already exists for the same memory cell, with a different symbol.
In other words, dynamic capabilities do not allow mutable may-alias relations, as those may compromise the alias tracking mechanism.

%% file: conclusion.tex
\section{Conclusion}

We introduced Fuel, a framework that can be plugged into existing compilers to extend their language's type system with capabilities and check for memory safety properties.
At the centre of the framework is a low-level language that describes the flow of capabilities in a program.
Through a series of examples, we showed how existing programming languages can be translated to Fuel's language and use the framework to ensure memory safety.

In its current state, our framework suffers from some limitations preventing its use in a concrete setting.
One is a lack of support of to represent self-recursive structures, such as linked-lists or trees, which would require our language to be augmented with recursive definitions and optional types.
Another relates to arrays, which would require the ability to compute indices dynamically.
Finally, Fuel does not yet support any form of concurrency.
Future work should aim at solving those issues, allowing the framework to be used in an existing language implementation's pipeline.